\documentclass[pdflatex,sn-nature]{sn-jnl}

\usepackage{graphicx}%
\usepackage{multirow}%
\usepackage{amsmath,amssymb,amsfonts}%
\usepackage{amsthm}%
\usepackage{mathrsfs}%
\usepackage[title]{appendix}%
\usepackage{xcolor}%
\usepackage{textcomp}%
\usepackage{manyfoot}%
\usepackage{booktabs}%
\usepackage{algorithm}%
\usepackage{algorithmicx}%
\usepackage{algpseudocode}%
\usepackage{listings}%
\usepackage{textcomp}
\usepackage{ragged2e}
\usepackage{setspace}
\newcommand{\Ca}{$^{40}$Ca$^+$}

\theoremstyle{thmstyleone}%
\theoremstyle{thmstyletwo}%

\theoremstyle{thmstylethree}%

\begin{document}

\title[Article Title]{\textbf{Scalable Trapped Ion Addressing with Adjoint-optimized Multimode Photonic Circuits}}


\author[1]{\fnm{Melika} \sur{Momenzadeh}}

\author[2]{\fnm{Ke} \sur{Sun}}

\author[2]{\fnm{Qiming} \sur{Wu}}

\author[2]{\fnm{Bingran} \sur{You}}

\author[2]{\fnm{Yu-Lung} \sur{Tang}}

\author[2,3,4]{\fnm{Hartmut} \sur{H\"affner}}

\author*[1,5]{\fnm{Maxim R.} \sur{Shcherbakov}}\email{maxim.shcherbakov@uci.edu}

\affil[1]{\orgdiv{Department of Electrical Engineering and Computer Science}, \orgname{University of California}, \orgaddress{
\city{Irvine}, \postcode{92697}, \state{California}, \country{USA}}}

\affil[2]{\orgdiv{Department of Physics}, \orgname{University of California}, \orgaddress{
\city{Berkeley}, \postcode{94720}, \state{California}, \country{USA}}}

\affil[3]{\orgdiv{Challenge Institute for Quantum Computation}, \orgname{University of California}, \orgaddress{
\city{Berkeley}, \postcode{94720}, \state{California}, \country{USA}}}

\affil[4]{\orgdiv{Computational Research Division}, \orgname{Lawrence Berkeley National Laboratory}, \orgaddress{
\city{Berkeley}, \postcode{94720}, \state{California}, \country{USA}}}

\affil[5]{\orgdiv{Department of Materials Science and Engineering}, \orgname{University of California}, \orgaddress{
\city{Irvine}, \postcode{92697}, \state{California}, \country{USA}}}


\abstract{Trapped-ion quantum computing requires precise optical control for individual qubit manipulation. However, conventional free-space optics face challenges in alignment stability and scalability as the number of qubits increases. Integrated photonics offers a promising alternative, providing miniaturized optical systems on a chip. Here, we propose a design for a multimode photonic circuit integrated with a surface-electrode ion trap capable of targeted and reconfigurable light delivery. 
Three closely positioned ions can be addressed using a focusing grating coupler that emits multimode light through electrode openings to 
ions trapped 80\,\textmu{}m above the chip. 
Simulations show that the couplers achieve diffraction-limited spot with a 4.3\,\textmu{}m beam waist along the trap axis and 2.2\,\textmu{}m perpendicular to the trap axis. Controlled interference of the TE$_{\text{10}}$ and TE$_{\text{20}}$ modes results in crosstalk of --20\,dB to --30\,dB at ion separations of 5-8\,\textmu{}m when addressing ions individually, and down to --60\,dB when two of the three ions are addressed simultaneously. Additionally, the higher-order TE modes can offer a novel mechanism for driving spin-motion coupling transitions, potentially enabling alternative approaches to quantum gates and simulations.
The proposed integrated platform offers a viable path for constructing large-scale trapped-ion systems, leveraging the benefits of nanophotonic design for precise and reliable ion manipulation.
}


\maketitle
\newpage 
\section{Introduction}\label{sec1}

Quantum computers have the potential to solve computational problems that are beyond the capabilities of classical computers.\cite{Feynman1982, Nielson2010, Preskill2018,horodecki2009} and hold transformative potential in  cryptography \cite{Shor1994}, complex optimization \cite{gutmann2014}, material simulation \cite{material2005}, and drug discovery \cite{Aspuru2005, Biamonte2017, montanaro2016}. However, a scalable, low-error platform for quantum computing is yet to be established. Among the various quantum computing platforms, trapped ions are particularly distinguished for their ability to meet all of DiVincenzo’s criteria \cite{Divincenzo2001}. Trapped ions offer advantages in long coherence times, high gate fidelity, and all-to-all entanglement mediated by shared motional modes arising from Coulomb interactions. \cite{Chiaverini2003,Chiaverini2005,Chiaverini2014,Karen2016,Home2020,Sandia2021,chiaverini2019}. 
Despite the promising potential of trapped ions for quantum computing, achieving scalability and portability remains a significant challenge in realizing a fully operational and practical quantum computer \cite{Wineland2008,Monroe2013}. 

For decades, free-space optics has been integral to ionization, cooling, state preparation, and the execution of coherent operations in trapped-ion systems \cite{Blatt2012,Cirac1995}. However, scaling these systems becomes increasingly impractical due to the large number of optical components required, as well as the limited optical access and alignment precision \cite{Kim2007, chiaverini2019, Moehing2007, Harty2014}, because the challenges of maintaining stable optical paths and achieving reliable qubit control grow rapidly as the number of ions increases \cite{Monroe2010, Lekitsch2017}.

Integrated photonics offer a compact
and scalable light delivery solution by combining optical waveguides, gratings, and modulators directly on a chip ~\cite{miller2009,Sun2015,shi2022,Jian-JunHe2017}. The feasibility of using integrated photonic devices to deliver light to ion traps has been shown with minimal alignment requirements ~\cite{Home2020,Sandia2021,Mouradian2022,Karen2016,karen2019grating}.

Efficient coupling of light into and out of thin optical waveguides remains a key challenge due to the moderate index contrast and limited radiation strength. Symmetric gratings are fundamentally constrained to a $3$\,dB coupling limit, as power is equally split between upward and downward diffraction. The coupling efficiency depends on the grating’s scattering strength ($\eta_1$), overlap with the fiber mode ($\eta_2$), and directionality ($\eta_3$). Apodized gratings can enhance $\eta_1$ and $\eta_2$ by spatially tailoring the scattering profile, while achieving high $\eta_3$ typically requires back-reflectors or structural refinement \cite{cheng2020,chen2010}. These improvements reduce higher-order diffraction and beam divergence, enhancing light delivery to ion qubits \cite{chen2016,cheng2020}. Designing such gratings requires optimization algorithms capable of handling complex, high-dimensional parameter spaces. While traditional methods like parameter sweeps and metaheuristics are computationally intensive \cite{J.Andkjær2010,J.Covey2013,W.S.Zaoui2014,Q.Zhong2014,B.Wohlfeil2014,R.Shi2014}, gradient-based optimization (GO) offers a scalable alternative. GO efficiently converges to optimal designs using only two simulations per iteration, making it well suited for inverse design of nanophotonic components \cite{yablanovich2013,Loncar2016,Vučković2017,sun2023}.

In this work, we propose the design of a photonic integrated circuit (PIC) for precise optical addressing of each of the three closely grouped calcium ions. Our design achieves individual ion addressing enabled by the interference of the TE$_{\text{10}}$ and TE$_{\text{20}}$ modes, with a cross-talk of $-21.1$\,dB at a lateral distance of $5$\,\textmu{}m from the center ion and $-31.5$\,dB at $8$\,\textmu{}m. To address two neighboring ions with the TE$_{\text{20}}$ mode, the cross-talk to the center ion is reduced to $-60$\,dB. The compact footprint of the PIC, measuring only $16 \times 45$\,\textmu{}m$^2$, enhances the scalability of the system. Additionally, we achieve a focusing efficiency (FE) of $-3.8$\,dB, corresponding to the optical power effectively delivered to the trapped ions at the trapping height. The proposed grating coupler, combined with on-chip phase and intensity modulation using integrated optics, provides a powerful platform for the scalable manipulation of trapped ions. 

\begin{figure*}[ht]
    \begin{minipage}{1\linewidth}
        \centering
        \includegraphics[width=\linewidth]{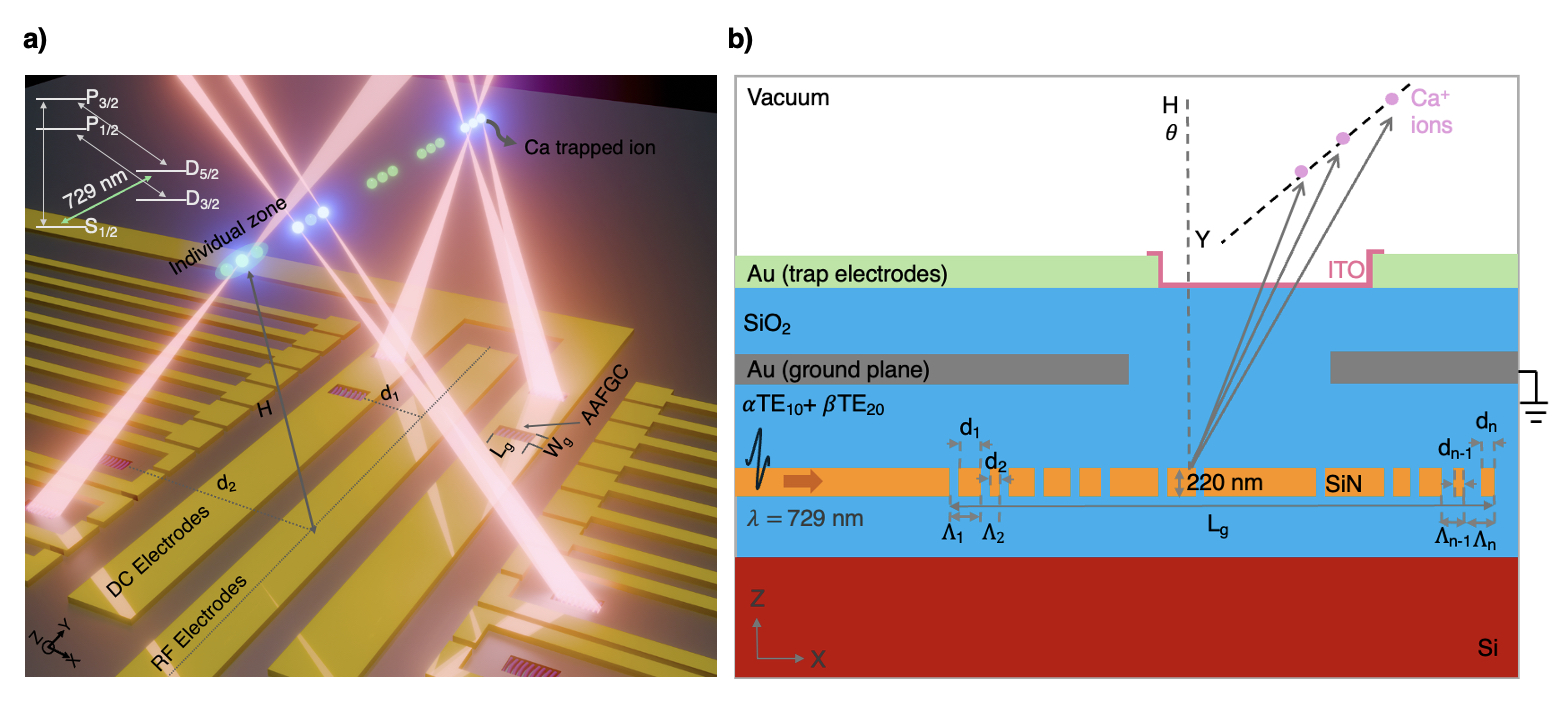}
        \justifying
        \caption{(a) Conceptual illustration of the designed surface trap with ions, integrated waveguides and beam forming elements underneath at multiple zones. Ions are trapped at one of the positions marked by the green dots, H= 80\,\textmu{}m above the electrodes, with appropriate potentials applied to the DC and radio-frequency (RF) electrodes.  The distances from the grating to the centerline of the trapping region are denoted as $d_1$ = 30\,\textmu{}m and $d_2$ = 100\,\textmu{}m. Top-left shows the energy-level diagram for \Ca. (b) Material stack of final ion trap devices, showing the integrated multimode waveguides and beam forming element beneath the surface-trap electrodes and light diffracting out of the chip by means of an apodized grating. Indium tin oxide (ITO) protects the ions from interface/substrate charging at the cost of reduced out coupling efficiencies.}
        \label{Concept}
    \end{minipage}
\end{figure*}

\section{Results}\label{sec2}
\textbf{Concept and Layer Stack}. Figure~\ref{Concept}(a) provides a schematic overview of the planar trap electrodes on the surface, with waveguides and beam-forming elements located $9.5$\,\textmu{}m beneath the top surface. Among the candidate ions, we selected calcium ions (\Ca) due to their well-characterized energy levels, efficient laser cooling, and compatibility with accessible laser wavelengths \cite{Home2020,Mouradian2022,Blatt2004,Innsburck2021,Innsburck2022}. Figure~\ref{Concept}(a) top-left includes energy level diagram for the \Ca\,ion, indicating the 729\,nm wavelength used for coherent qubit operations. The ions are trapped $80$\,\textmu{}m above the surface in five distinct zones, with each zone containing three ions. To accommodate fabrication tolerances and optimize trapping performance, multiple grating variations have been positioned at different distances from the trap centerline ($d_1$ and $d_2$). The device cross section in Figure~\ref{Concept}(b) illustrates forward-emitting diffraction to three \Ca\,ions trapped along Y-axis. Single ion addressing is achieved by carefully selecting the phase and intensity of TE$_{\text{10}}$ and TE$_{\text{20}}$ modes using an asymmetric mode converter and multi-mode interferometer (MMI) with \textmu{}s switching speed \cite{pradip2024fast,chung2019low,yong2022power} to create appropriate out-coupled patterns. The top gold (Au) layer represents trap electrodes that generate the electric field required for ion confinement and the second Au layer is used as a ground plane to shield against unwanted noise. Oxide layers with thicknesses of $5$\,\textmu{}m provide optical confinement. We chose silicon nitride (SiN) as the platform due to its low propagation loss, wide transparency window, and compatibility with complementary metal-oxide semiconductor fabrication processes.

\begin{figure*}[ht]
    \centering
    \includegraphics[width=0.95\linewidth]{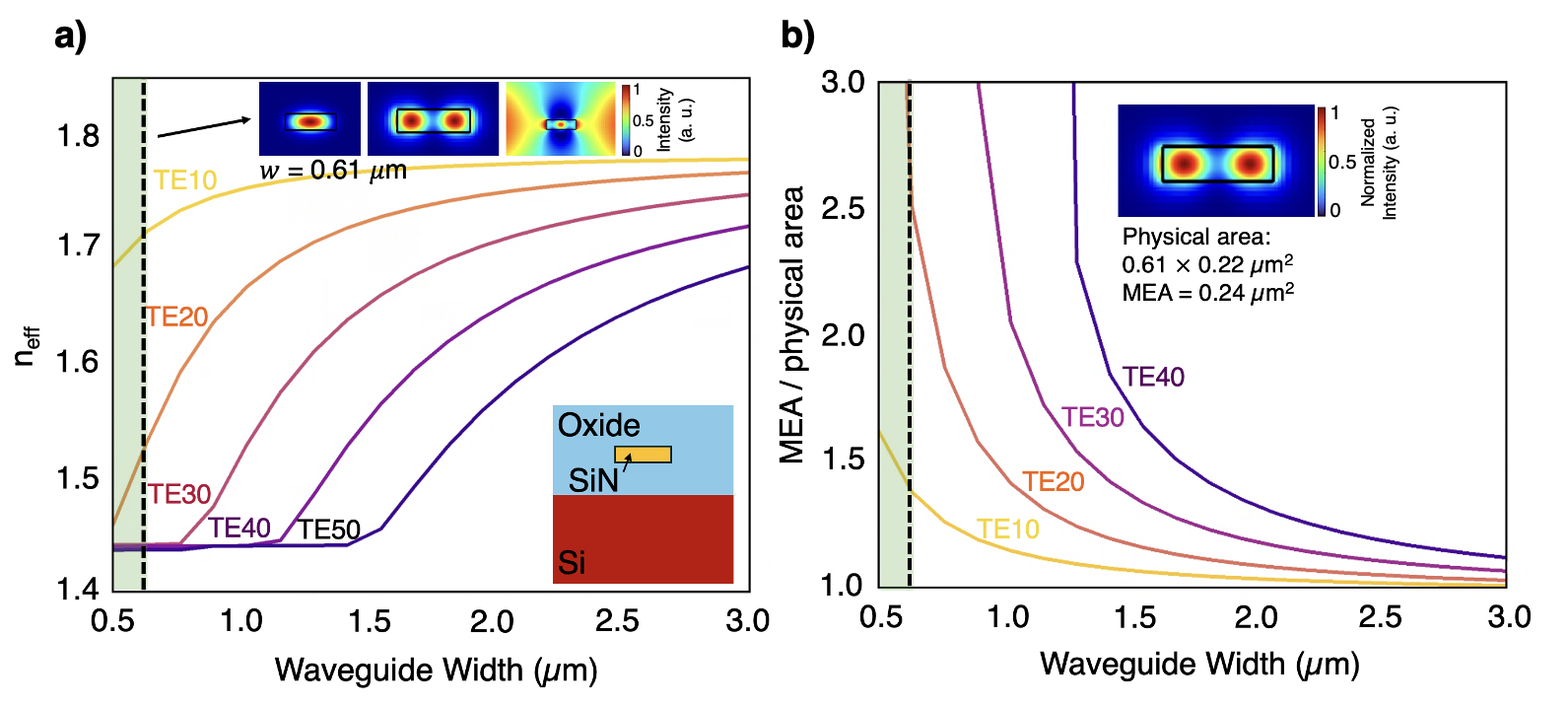}
    \caption{(a) Effective refractive index $(n_{\text{eff}})$ and (b) Normalized mode effective area (MEA) as a function of the waveguide width for different TE modes. The shaded regions indicate the optimal width range for efficiently supporting TE$_{\text{10}}$ and TE$_{\text{20}}$ modes while cutting off higher-order modes.}
    \label{Waveguide}
\end{figure*} 

\textbf{Optical Waveguide Design.} The effective refractive index $(n_{\text{eff}})$ and propagation constant (\(\beta\)) in waveguides are distinct for each mode:
\begin{equation}
\beta_{m,n} = \frac{2\pi}{\lambda}n_{\text{eff}(m,n)},
\label{eq:beta}
\end{equation}
where $m$ and $n$ are the mode indices. For a mode to be confined within the waveguide, it must satisfy the condition for total internal reflection $k_{\text{0}} n_{\text{clad}} < \beta < k_{\text{0}} n_{\text{core}}$, where $n_{\text{clad}}$ and $n_{\text{core}}$ are the refractive index of the cladding and core materials, respectively. The width of the waveguide is chosen to ensure support for the TE$_{\text{10}}$ and TE$_{\text{20}}$ modes, while cutting off higher-order modes with a free-space wavenumber of $k_{\text{0}} = \frac{2\pi}{\lambda}  = 8.62$~$\mu$m$^{-1}$. Figure~\ref{Waveguide}(a) shows the effective index of the guided modes, where higher-order modes emerge at larger waveguide widths. Insets depict the mode profiles for a width of $w$, highlighting the transition from weak confinement for the TE$_{\text{30}}$ to strong confinement for the TE$_{\text{10}}$ and TE$_{\text{20}}$. The green-shaded region highlights the optimal waveguide width range for mode selection confinement. The bottom-right inset illustrates the cross-sectional structure of the waveguide, where a SiN core (refractive index $n = 2.0$) is embedded in an oxide cladding (refractive index $n = 1.4$) atop the silicon substrate. Figure~\ref{Waveguide}(b) highlights the relationship between the waveguide width and the mode effective area (MEA), demonstrating how the waveguide dimensions affect the supported modes' areas and the cutoff for higher-order modes. The green shaded region indicates a waveguide width that supports the TE$_{\text{10}}$ and TE$_{\text{20}}$ modes, while effectively suppressing higher-order modes due to their sharp transition into radiation modes and reduced confinement within the physical area of the waveguide. The numerical modal analysis of the waveguide, based on the propagation constant $\beta_{\text{m,n}}$, ensures double-mode operation within the defined width range of $w <$ $0.61$\,\textmu{}m. The cross-sectional area of the designed waveguide is $0.22$ \(\times\) $0.6$\,\textmu{}m$^2$. The thickness has been chosen to conform to common foundry standards.
\begin{figure*}[ht]
    \centering
    \includegraphics[width=0.95\linewidth]{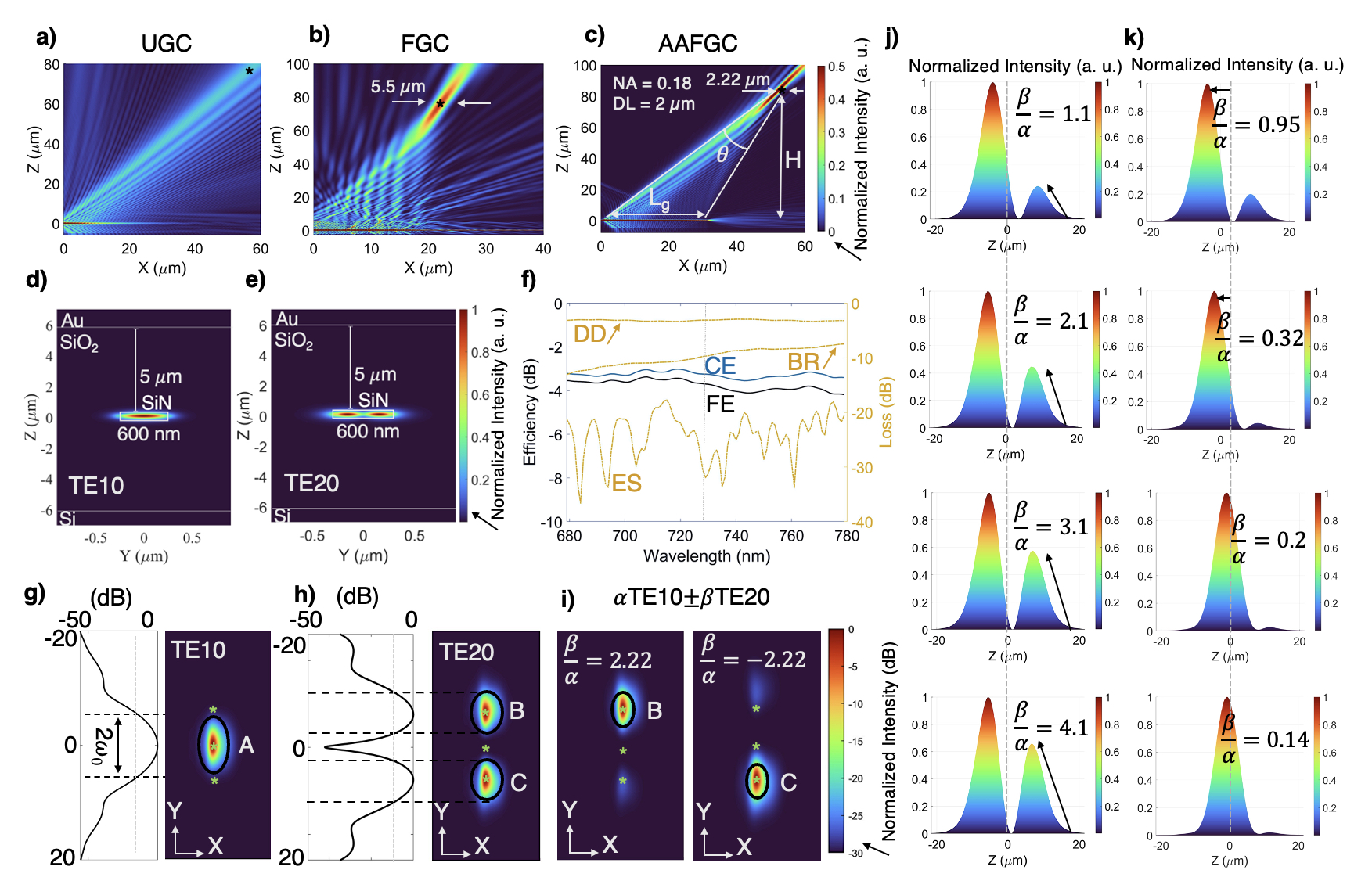}
    \caption{(a) The uniform grating coupler  (UGC) demonstrates a broad diffraction profile with weak focusing. (b) The focusing grating coupler (FGC)  enables enhanced focusing with a focal spot size = 5.5\,\textmu{}m. (c) Adjoint-based apodized focusing grating coupler (AAFGC) 
    features an emission angle of 40\textdegree{}, a numerical aperture (NA) of 0.18, and focusing to a diffraction-limited spot of 2.22\,\textmu{}m; the diffraction limit (DL) is 2\,\textmu{}m. 
    (d-e) Waveguide cross sections displaying the quasi-TE mode profiles for the TE$_{\text{10}}$ and TE$_{\text{20}}$ modes in the proposed material stack. (f) Efficiency and loss analysis of the AAFGC. The panel shows end-scattering (ES), back reflection (BR), downward diffraction (DD) losses, coupling efficiency (CE), and focusing efficiency (FE). (g-i) Intensity patterns at the trap height (z = $80$\,$\mu$m) along the X and Y directions for the TE$_{\text{10}}$ and TE$_{\text{20}}$ modes and their combinations with $\frac{\beta}{\alpha}$ = 2.22 to address Ion B and -2.22 to address Ion C. (j,k) Intensity profile for various $\frac{\beta}{\alpha}$ ratios.}
    \label{Results}
\end{figure*}

\textbf{Adjoint-based Apodized Focusing Grating Coupler (AAFGC) Design.} To enable efficient and flexible multi-ion addressing, we design a beam-forming element optimized to suppress cross-talk, maximize focusing efficiency, and minimize the beam spot size. By dynamically mixing two  waveguide modes
, the grating enables precise beam shaping for selective addressing of three ions in reconfigurable spatial configurations. We started by designing a two-dimensional (2D) uniform grating coupler (UGC) as shown in Figure~\ref{Results}(a) using a longitudinal 
approach \cite{karen2019grating} to achieve positive emission angles, focusing on the 2D cross-section along the grating centerline (X-Z plane). 
Emission characteristics, including the emission angle (\(\theta\)) and grating strength (\(\xi\)), are primarily influenced by the grating period (\(\Lambda\)) and the duty cycle (DC), which defines the ratio of the SiN-filled region to the total period; see Supplementary Information Figure~\ref{fig:initial_params}(c-d). The grating strength \(\xi\) quantifies the exponential decay of the guided mode amplitude along the grating, expressed as $I(x) = I_0$\(e^{-\xi x}\). The allowable range for the grating period is governed by the Bragg condition, ensuring efficient out-coupling and providing the desired emission angle. This condition is given by:
\begin{equation}
n_\text{eff} - \frac{m \lambda}{\Lambda}
= n_\text{clad}\sin\theta,
\label{eq:2}
\end{equation} 
where $n_{\text{eff}}$ is the effective refractive approximated by a linear interpolation as $n_\text{eff} \approx DC n_\text{clad}+(1-DC)n_\text{core}$, \(\theta\) is the emission angle, and $m=1$ is the diffraction order. We investigated the TE$_{\text{10}}$ mode at $\lambda=729$\,nm, suitable for the coherent control of \Ca optical qubits, with SiO$_{\text{2}}$ as the cladding material \cite{Fraser2024, Vitali2022}. Fabrication constraints were imposed by a minimum feature size of 200\,nm and a minimum spacing of 100\,nm, as imposed by foundry fabrication processes.
The grating strength demonstrates the impact of substrate and  oxide-vacuum interface reflections, heavily influenced by the top and bottom oxide layer with thicknesses of  (5\,\textmu{}m).

To surpass the performance of the UGC, a particle swarm algorithm is employed for the coarse initialization of key design parameters. For a linearly apodized grating, the DC is spatially varied along the waveguide as: $DC(x) = DC_\text{0} - Rx,$
where $DC_0$ denotes the initial duty cycle and R is the apodization rate. $DC_0 = 1$ corresponds to a fully solid waveguide; we used a reduced value of $DC_0 = 0.95$ to mitigate the fabrication challenges associated with subcritical trench dimensions. Figure~\ref{Results}(b) shows the designed FGC with a beam spot size of 5.5\,\textmu{}m and a focusing efficiency of –4.5\,dB. This performance is limited by the coarse optimization capabilities of the particle swarm algorithm, which struggles to effectively explore high-dimensional parameter spaces. Achieving further improvements in beam confinement and efficiency necessitates the use of inverse design techniques, such as adjoint-based optimization.

To further enhance the performance of the FGC, GO was employed, enabling efficient exploration of the high-dimensional design space \cite{sun2023}. 
The optimization process was conducted under realistic fabrication constraints, including fixed etch depth, layer thickness tolerances, and a minimum feature size of 200\,nm. The AAFGC exhibits improved beam focusing, reduced side lobes, and enhanced radiation efficiency. The results of this optimization are shown in Figure~\ref{Results}(c), highlighting the spatial profile of the output field and the effectiveness of the tailored apodization. Extending from the 2D design, we introduced curvatures to the grating lines to achieve three-dimensional (3D) control over the effective refractive index ($n_{\text{eff}}$) and phase. 
The final design, incorporating a mode-expanding taper and a series of curved grating lines with varying periods, duty cycles, and radii of curvature, optimized coupling efficiency and ensured accurate light delivery near the ion location 
Figure~\ref{Results}(f) demonstrates the simulated efficiency and losses of the beamforming element which include coupling efficiency (CE), focusing efficiency (FE), and losses contain end scattering (ES) 
, downward diffraction (DD) and back reflection (BR) loss to the input port. 
Figure~\ref{Results}(g-i) shows the resulting intensity profiles of the light emitted along the X and Y directions for the four configurations. The emitting region of each coupler for the fundamental mode has an area of $2.22 \times 4.25$\,\textmu{}m\(^2\). 
A comprehensive AAFGC design overview is provided in the Methods section.  

\textbf{Ion Addressing.} Consider a scenario where three ions are trapped $80$\,\textmu{}m above the photonic chip. We explore four distinct ion addressing configurations to implement targeted light delivery for quantum operations. Configuration 1 involves selectively addressing the center ion (Ion A); configuration 2 targets the two outside ions (Ions B and C) simultaneously; configuration 3 focuses on Ion B, while configuration 4 addresses Ion C. To accomplish this, we manipulate the beam profiles by employing different input modes and combining them as needed. For configuration 1, we use the TE$_{\text{10}}$ mode, which provides a well-centered and focused beam, ensuring exclusive light delivery to the Ion A. In configuration 2, addressing both outside ions simultaneously is achieved using the TE$_{\text{20}}$ mode, minimizing unintended exposure to the center ion. For configurations 3 and 4, we employ a linear combination of the TE$_{\text{10}}$ and TE$_{\text{20}}$ modes. The general form of this combination is expressed as:
\begin{equation}
E(x) = \alpha \mbox{TE}_{\text{10}} + \beta \mbox{TE}_{\text{20}},
\label{eq:3}
\end{equation} 
where \(\alpha\) and \(\beta\) are adjustable coefficients determining the relative contribution of each mode. By tuning the amplitude and phase of these modes, we can generate asymmetric beam profiles that direct the light to the desired ion (Figure~\ref{Results}(j-k)). This mode engineering strategy provides a robust and adaptive solution for multi-ion addressing in trapped ion quantum systems.

\begin{figure}[ht]
\centering
\includegraphics[width=0.95\linewidth]{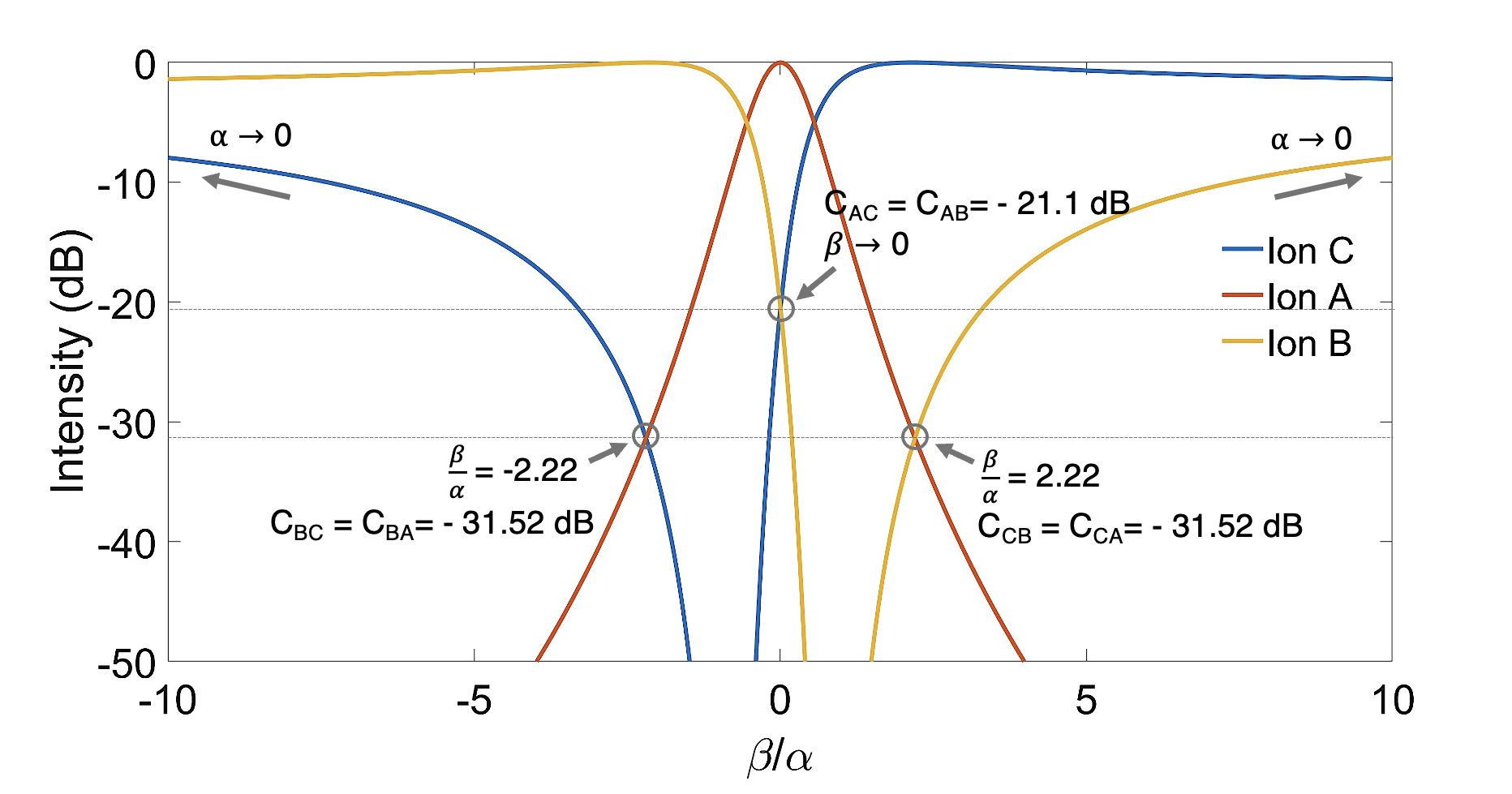}
\caption{Intensity at Ion A, B, and C sites as a function of $\frac{\beta}{\alpha}$ for ions at 5\,\textmu{}m from each other. The plot illustrates the capabilities of the chip to minimize crosstalk between the ions, showing the crosstalk coefficients ($C_{\text{ij}}$) for various $\frac{\beta}{\alpha}$ values.
}
\label{fig:crosstalk}
\end{figure}

\textbf{Crosstalk Characterization.} Crosstalk suppression is crucial in trapped-ion systems as it directly impacts the fidelity of quantum operations \cite{Oxford2014,Wineland2008,Monroe2013,Figgatt2019}. Figure~\ref{fig:crosstalk} illustrates the optical intensity at the positions of three ions (A, B, and C) as a function of $\frac{\beta}{\alpha}$. By tuning $\frac{\beta}{\alpha}$, the optical focus shifts along the ion chain, enabling selective excitation of individual ions while minimizing light delivery to the neighboring sites. The figure demonstrates strong localization with crosstalk suppression, reducing unwanted signal to approximately $C_{\text{BC}}= C_{\text{BA}}=-31.5$\,dB for non-targeted ions at $\frac{\beta}{\alpha} = -2.22$, where ion B is selectively addressed. At $\beta = 0$, only the fundamental mode is excited, resulting in the focusing on Ion A with crosstalk levels of $C_{\text{AC}} = C_{\text{AB}} = -21.1$\,dB. As $\alpha$ approaches 0, Ions B and C are addressed, with the intensity at Ion A minimized, leading to crosstalk levels below $C_{\text{BA}}=C_{\text{CA}}=-60$\,dB. The achieved levels are generally suitable for many quantum information processing tasks \cite{crosstalk1,crosstalk2,crosstalk3}.

\textbf{Pseudo-potential simulation}.
In ion traps, ions are typically confined by oscillating RF fields in two radial directions and static fields in the axial direction~\cite{paul1990electromagnetic}. The equilibrium position is located at the minimum of the combined potential formed by the static field from the DC electrodes and the pseudo-potential generated by the RF field. As a result, any imperfections in the trapping electrodes can affect the ion trap's performance.  
Previous studies have demonstrated various types of surface-electrode traps with solid performance~\cite{revelle2020phoenix,noel2019electric},
including grating couplers
~\cite{Chiaverini2020,Home2020}. These couplers, positioned beneath the dc or RF electrodes, can introduce perturbations to the ion confinement potentials. 

\begin{figure}[ht]
\centering
\includegraphics[width=0.75\linewidth]{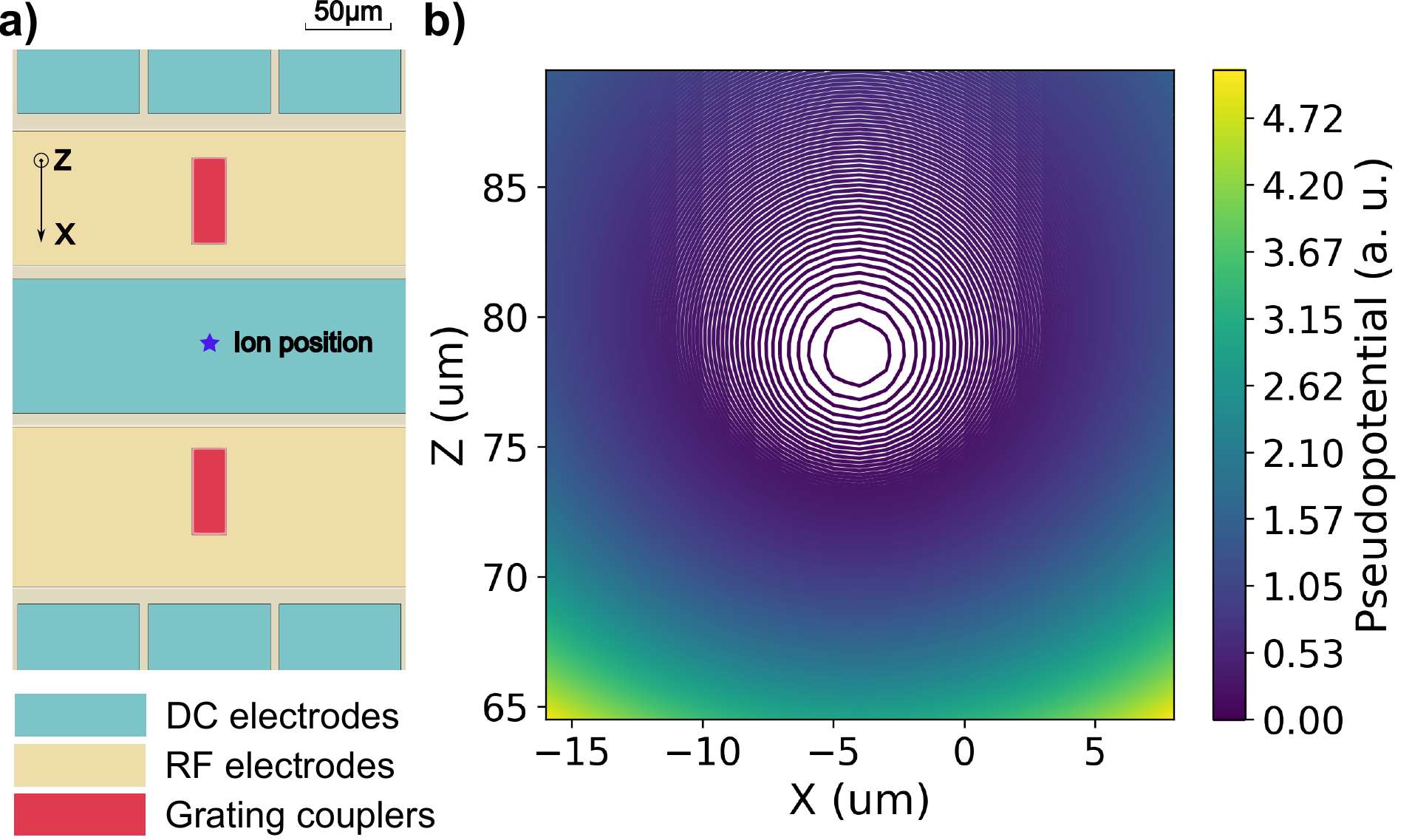}
\caption{\textbf{(a)} A section of the trap's layout. \textbf{(b)} The pseudopotential generated by the RF electrodes at the location of the ion (star in (a)). X and Z represent directions parallel and perpendicular to the trap surface, respectively. The color gradient illustrates the potential, revealing a minimum near $(X,Z) = (-4.5, 80)$\,\textmu{}m. The displacement along the X direction mostly results from the intentionally asymmetric design of the RF electrodes.}
\label{fig:trap}
\end{figure}

In our design, the grating couplers positioned beneath the RF electrodes (Fig.~\ref{fig:trap} (a)), necessitating an opening in the center of the long RF rail to allow light transmission. To evaluate the impact of this opening, we use the boundary element method (BEM; see Methods) to analyze its effect and compare it with regions without grating couplers. Our results, shown in Fig.~\ref{fig:trap} (b),  indicate that the predicted trapping height remains approximately 80\,\textmu{}m, with a difference of less than 0.5\,\textmu{}m between the two cases, suggesting a negligible perturbation of trapping performance. 

\section{Discussion}
This study proposes a design for integrating photonic devices into a surface ion trap, allowing for the delivery of light to each ion in a mutiple-ion chain. This capability offers significant advantages for the trapped-ion technology, while the integrated approach enhances stability and reproducibility. Below, we outline some potential applications of our design. Importantly, by increasing the number of coupled waveguide modes TE$_{m,n}$ and TM$_{m,n}$, the number of addressable ions can be scaled up accordingly.

\textbf{Individual measurement.}
Measurement is typically the final step in a quantum circuit, providing the final quantum states of the qubits, which constitute the quantum information. Recent advances have developed strategies for mid-circuit measurements~\cite{motlakunta2024preserving,pino2021demonstration}, which are particularly valuable for simulating measurement-induced quantum phase transitions \cite{noel2022measurement, czischek2021simulating, lavasani2021measurement} and conducting quantum error corrections \cite{ryan2021realization}, which require the measurement of ancilla qubits during the execution of the circuit. A recent study~\cite{motlakunta2024preserving} demonstrated the capability to perform individual measurements on an ancillary ion by tightly focusing the detection laser beam. The study reported minimal interference with the spin state of neighboring data ions during the measurement, thereby highlighting the feasibility of mid-circuit measurements. In contrast to the free-space optics employed in \cite{motlakunta2024preserving}, integrated photonic devices offer superior laser stability, as they are co-located on the same chip as the ion trap, thereby sharing common mechanical and thermal fluctuations and providing exceptional laser coherence. Moreover, integrated photonic devices facilitate system scaling to multiple distant zones, where a single high-NA lens would struggle to cover all regions effectively.

\textbf{Individual gates.}
Individually implemented gates are particularly crucial for long ion chains, where multiple ions are confined within the same trapping potential and ion shuttling is infrequent \cite{monroe2021programmable}. In such scenarios, individual control of specific ions relies on tightly focused laser beams targeting the desired ions. Various techniques have been developed to achieve individual addressing, including multi-channel acousto-optic modulators \cite{debnath2016demonstration}, microelectromechanical systems \cite{wang2020high}, and acousto-optic deflectors \cite{pogorelov2021compact}. These methods have demonstrated high-fidelity qubit operations and effective individual control. However, these addressing systems are typically bulky and assembled outside the vacuum chamber. Mechanical fluctuations along the beam path can lead to qubit decoherence. In contrast, our approach offers inherent stability and the potential to enhance qubit coherence times.

\textbf{Ion-TE modes coupling.}
The operation of trapped-ion systems is primarily based on laser-ion interactions, specifically the interaction between the electric field of the laser beam and the dipole or quadrupole moment of the ions. Current studies normally utilize Gaussian beams to drive ion transitions, assuming a uniform electric field at the ion's position under ideal beam alignment. In this scenario, the sideband transition, which is critical for implementing two-qubit gates and motion-related quantum simulations~\cite{debnath2016demonstration, sun2024quantum}, is generated by the laser interacting with the first order of Taylor expansion of the ion's position operator. Control over whether the laser couples to the zero-order term (carrier transition) or the first-order term (sideband transition) is achieved by tuning the laser frequency to the corresponding resonance. However, in this case, the unwanted off-resonant coupling to carrier transitions is unavoidable when driving sidebands, introducing errors in quantum operations~\cite{Karen2024}. Moreover, the strength of the sideband coupling is suppressed by the Lamb-Dicke parameter~\cite{wineland1998experimental}, typically around 0.1. This strength will be further reduced when dealing with higher-order sideband transitions, which are preferred in quantum simulations~\cite{kang2024seeking} and multi-qubit operations~\cite{katz2023programmable}. 

Recent studies have proposed leveraging higher-order TE modes as an alternative method for coupling the laser field to the motional modes of ions, potentially circumventing the coupling limitations imposed by the Lamb-Dicke parameter and providing carrier-free sideband operations~\cite{Karen2024}. The position-dependent electric field gradients in higher TE modes enable coupling to the ion's motion, introducing a new scaling parameter $\eta_{0n}\sim nx_0/a$, which is governed by the position gradient of the electric field $1/a$ and TE modes. With engineered mode choices, this factor can approach the order of unity $\eta_{0n}\sim1$. In this context, the integrated photonics approach proposed here is highly suitable. By configuring MMIs, it becomes possible to switch between the TE$_{\text{10}}$ and TE$_{\text{20}}$ modes, allowing precise coupling to different orders of the ion's motional modes.

\textbf{Conclusion.}
In summary, we designed and optimized integrated photonic components for selective light delivery to three close \Ca  ~ions, employing phase control for beamforming to generate tailored light patterns at the ion sites. Through adjoint optimization, we developed focusing grating couplers with a focusing efficiency exceeding $-4$\,dB, while maintaining an insertion loss below $2.7$\,dB across the operational wavelengths. The beamforming achieved diffraction-limited spots, with crosstalk levels ranging from $-20$\,dB to $-30$\,dB when exciting the neighboring ions using the fundamental mode and $-60$\,dB when the two outer ions are addressed simultaneously.
The proposed mode-based beamforming approach offers a versatile and reconfigurable solution for precise light delivery in various ion configurations without free-space optics, paving the way for scalable and efficient implementations in trapped-ion quantum technologies.

\section{Methods}
\textbf{Design procedure.} 

The mode properties (mode distribution and effective refractive index) of the guided modes in the SiN waveguides are calculated by using finite-element method (FEM), and optical simulations were performed using Lumerical FDTD (Ansys Inc.), with GO performed via Ansys Lumerical Python APIs which provide lumapi automation capabilities, and lumopt module. The design process is broken down into distinct steps:
\begin{itemize}
    \item Material Stack and Grating Structure Design:
The material stack and grating coupler structure are designed while considering fabrication constraints, including limitations on etch depth, minimum feature size, and layer thicknesses. These factors influence the achievable grating parameters and emission characteristics.
\item Look-Up Table Generation:
A comprehensive look-up table is generated by simulating the emission angle, decay constant, and effective index across a broad range of duty cycles (DC) and grating periods \(\Lambda\). The objective is to ensure a positive emission angle (\(\theta\) $> 0$) within the operational wavelength. Uniform gratings are used for this preliminary analysis, and the results are presented in Figure~\ref{fig:initial_params}.
\item Coarse initialization:
A linear apodization profile was applied to the uniform grating coupler (UGC) design, and coarse optimization using a particle swarm algorithm was employed to identify near-optimal initial parameters for subsequent adjoint-based refinement. 
\item Adjoint Optimization Initialization:
The initial parameters obtained from the step 3 serve as the starting point for the apodization of the FGC using adjoint optimization. Proper initialization is critical for the convergence of the optimization process, and poor initialization can lead to prolonged runtime and convergence failures.
\item Figure of Merit (FOM) Definition:
Multiple FOMs are established based on the design goals, which include minimizing optical crosstalk, maximizing the focusing efficiency at a specified focal position and height, and reducing the focused beam spot size close to the diffraction limit \(\lambda\)/$2NA$). A minimum of three FOMs is applied, each weighted according to its priority to balance the design trade-offs.
\item 2D Optimization:
To reduce computational overhead, the initial optimization is performed in two dimensions. This enables faster convergence while capturing key design characteristics.
\item 3D Optimization:
Following the 2D optimization, the design is expanded to 3D. The waveguide width is optimized to ensure support for the two target modes, TE$_{\text{10}}$ and TE$_{\text{20}}$, while suppressing higher-order modes. Additionally, the width of the AAFGC is further refined to achieve optimal focusing along Y based on the desired focal distance and beam spot size. Curved grating teeth are employed to enhance focusing efficiency (FE) by reducing mode mismatch and minimizing back-reflection losses. The optimized opening angle and taper length are presented in Figure~\ref{fig:opening_angle}.
\item Asymmetric Mode Converter:
A mode converter is employed to transition between a single-mode input and the multimode waveguide supporting TE$_{\text{10}}$ and TE$_{\text{20}}$. The converter is carefully optimized to minimize insertion losses and ensure high overlap with the desired modes, as illustrated in Figure~\ref{fig:ADC}, where the mode-dependent spatial shift is also analyzed.
\item Multimode Interferometer (MMI):
A multimode interferometer is integrated to enable coherent power mixing of the TE$_{\text{10}}$ and TE$_{\text{20}}$ modes. The MMI geometry is optimized for the equal-power distribution with minimal phase mismatch, providing robust mode control within the multimode section of the circuit. The simulated mode profile at the output plane confirms successful mode excitation, as shown in Figure~\ref{fig:MMI}.
\end{itemize}

\textbf{BEM Simulation.}
The properties of electrodes, including RF potentials and trapping stability, are analyzed using Boundary Element Method (BEM) simulations. The workflow consists of distinct stages:
\begin{itemize}
    \item Trap model creation (STL format):
The trap geometry is modeled using computer-aided design (CAD) software like Fusion 360, ensuring that it is in accordance with optical coordinate conventions (trap axis along the z-axis). The electrodes are distinctly colored using a predefined BEM-compatible color scheme to clearly differentiate the individual electrodes. The trap model is exported as a binary STL file with embedded electrode-identification attributes.
\item Mesh preparation in BEM:
The exported STL file is imported into BEM and re-meshed with refined mesh density to accurately resolve electrode features. Mesh integrity is confirmed through 3D visualization to ensure proper representation of electrode boundaries and surface discretization.
\item Electrostatic simulation:
Electrostatic simulations are performed using BEM to calculate the potential distributions generated by each individual electrode. For DC electrodes, unit voltages are applied sequentially to each electrode while grounding the others, enabling precise characterization of their individual contributions to the total trapping potential. For RF electrodes, the time-averaged pseudo-potential is computed to evaluate confinement in the radial directions. Simulations can be executed locally or on high performance computing cluster (Berkeley HPC, Savio), depending on computational complexity and desired resolution.
\item Result analysis:
The simulation outputs, including electrode potentials, electric field distributions, and surface charge densities, are visualized and analyzed in 3D. Further detailed multipole analysis and derivative evaluations, essential for trapping performance characterization, are conducted using the PaulTrapAnalysis Python package. This comprehensive evaluation ensures accurate predictions of trap characteristics, stability, and the impacts of photonic integration cut-outs.
\end{itemize}

\backmatter


\bmhead{Acknowledgements}
We acknowledge the assistance of J. Bocanegra and M. McMorris, and stimulating discussions with A. Jayich, M. Radulaski, and R. Kwapisz.

\bmhead{Funding}
This project was supported by the UC Noyce Initiative 2024 Quantum Information Science program.

\bmhead{Data availability}
Data available from the corresponding author upon reasonable request.




\newpage


\bibliography{References}
\newpage

\begin{appendices}

{\center \Large \textbf{Supplementary Information}}

\section{Beam Forming Element}\label{secA1}
\begin{figure*}[ht]
    \begin{minipage}{1\linewidth}
        \centering
        \includegraphics[width=\linewidth]{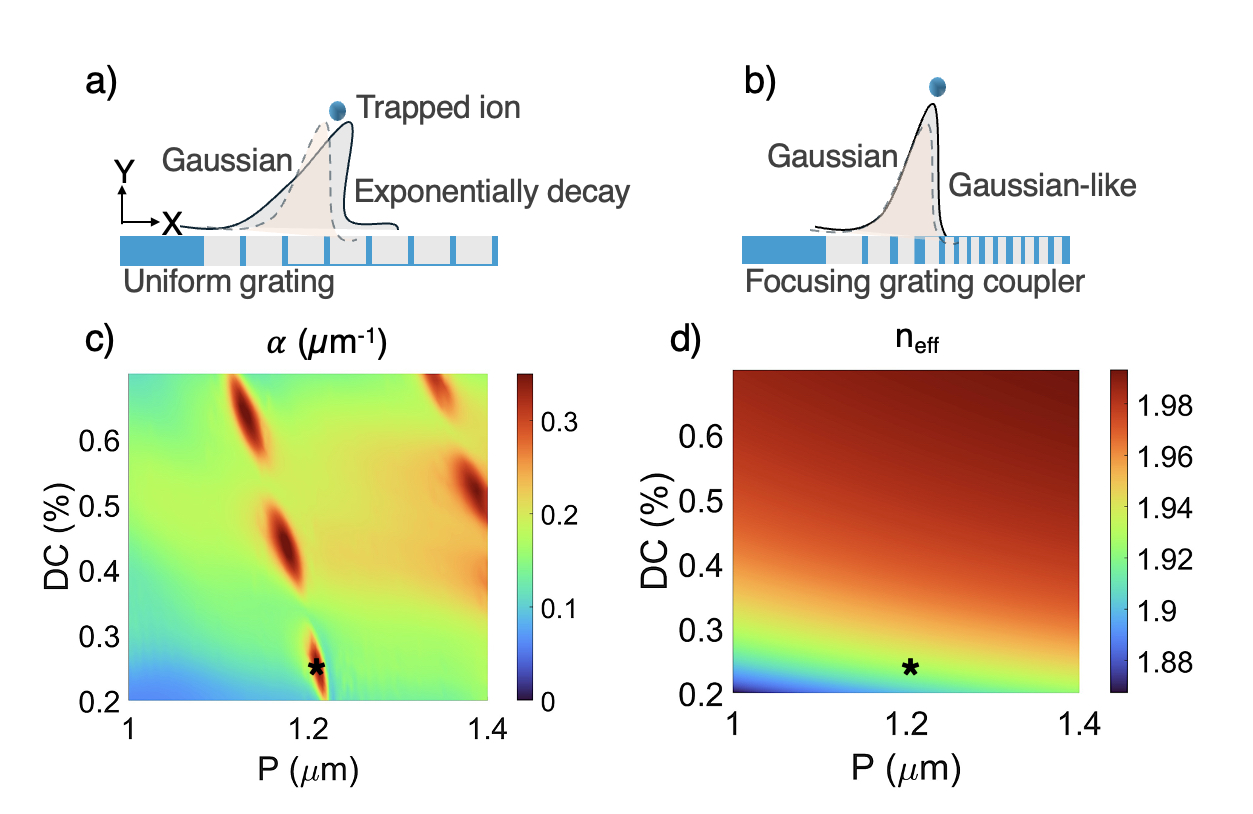}
        \justifying
        \caption{Comparison of light emission from a uniform grating coupler and a focusing grating coupler. (a) A uniform grating coupler generates a Gaussian beam with significant exponential decay, leading to inefficient light delivery to the trapped ion. (b) A focusing grating coupler with a non-uniform design produces a Gaussian-like beam with enhanced confinement and reduced diffraction losses, ensuring improved coupling efficiency for ion-trap applications. (c-d) Grating strength and effective refractive index of uniform grating coupler.}
        \label{fig:initial_params}
    \end{minipage}
\end{figure*}

Non-uniform grating couplers are designed to efficiently couple light from a waveguide into free space or a specific target, such as a trapped ion. Unlike uniform gratings, which have constant periodicity and coupling strength, apodized gratings feature a gradual variation in parameters like grating period, duty cycle, or etch depth. This tailored design allows better control over the emitted light’s phase and amplitude, producing a focused, Gaussian-like beam with reduced diffraction and minimized radiation loss. Additionally, focusing grating couplers can optimize mode matching, resulting in improved coupling efficiency, which is particularly beneficial for applications in quantum optics and trapped-ion systems.
In the Figure~\ref{fig:initial_params}(a), illustrates a uniform grating coupler, where the Gaussian-like input field results in an exponentially decaying output, leading to poor light delivery and inefficient coupling to the trapped ion. In contrast, Figure~\ref{fig:initial_params}(b) demonstrates a focusing grating coupler that generates a Gaussian-like beam with improved directional control and reduced divergence. This design enables more efficient and precise light delivery to the trapped ion, making it advantageous for quantum information processing. The focusing effect is achieved by the non-uniform grating structure, which ensures the emitted light remains well-confined and aligned with the target.
\begin{figure*}[ht]
    \begin{minipage}{1\linewidth}
        \centering
        \includegraphics[width=\linewidth]{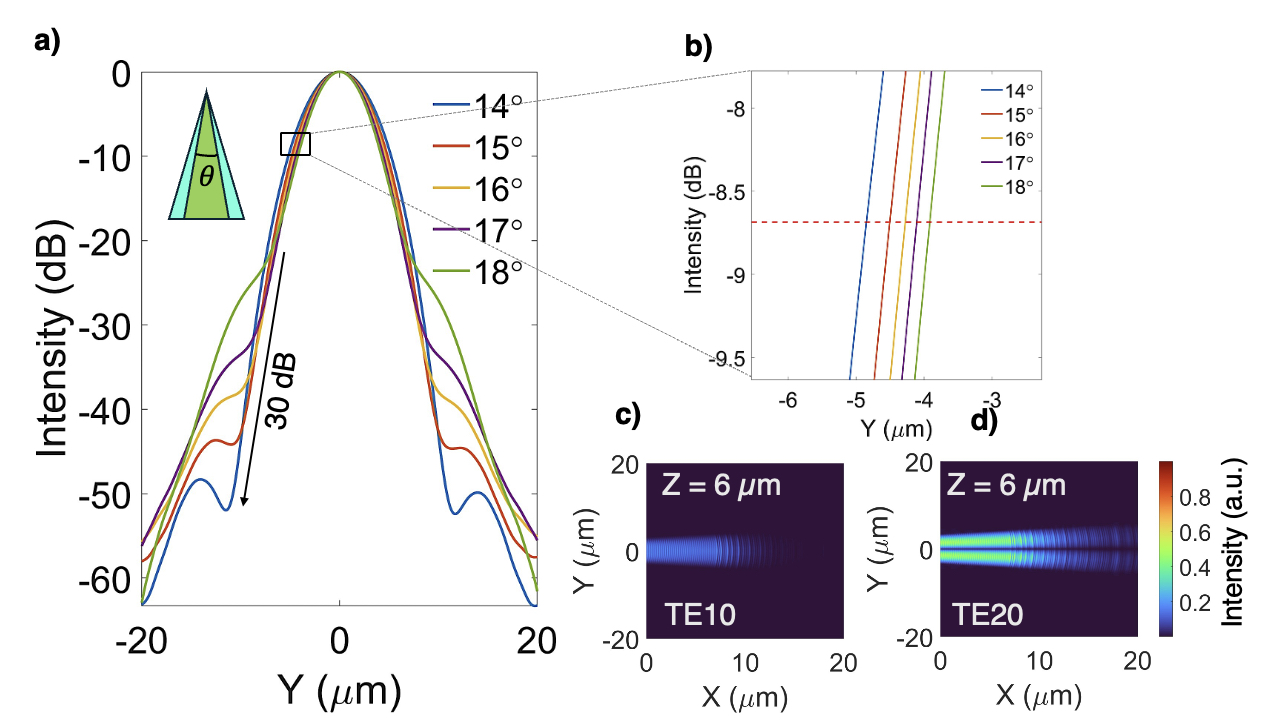}
        \justifying
        \caption{Apodized grating 3D results, (a) Cross-talk dependence to opening angle of grating coupler, (b) Beam-spot size relation to the opening angle, (c,d) Normalized intensity at the center of the grating for two modes.}
        \label{fig:opening_angle}
    \end{minipage}
\end{figure*}
To systematically explore the parameter space and evaluate performance improvements using gradient descent with the adjoint method, we performed a simple parameter sweep of uniform grating configurations. Specifically, we varied the grating period from 1 to 1.4\,\textmu{}m in increments of 0.025\,\textmu{}m and adjusted the duty cycle from 0.2 to 0.8 in steps of 0.01. This approach allowed us to capture a wide range of geometrical configurations and their impact on light transmission and focusing efficiency.  Figure~\ref{fig:initial_params}(c) illustrates grating strength, defined as the local coupling coefficient between the guided mode and radiated field. A constant grating length of 50\,\textmu{}m sufficed for accurate extraction of parameters. Figure~\ref{fig:initial_params}(d) illustrates the effective refractive index along the grating, highlighting the balance between optical confinement within the waveguide and out-coupling into free space. The maximum focusing efficiency recorded was -6.67\,dB, while the coupling efficiency, the total transmitted power scattered from the grating—reached -3.1\,dB at a period of 1.2 microns and a duty ratio of 0.25. The difference of 3.56\,dB between coupling and focusing efficiency indicates significant scattering losses that prevent uniform gratings from achieving optimal focusing. This result highlights the fundamental limitation of using a uniformly spaced grating, as a considerable portion of the transmitted power is scattered rather than focused. To address this limitation and achieve focusing efficiencies approaching the coupling limit, non-uniform grating structures optimized via gradient descent are necessary. By leveraging the adjoint method, it is possible to iteratively refine the grating geometry to minimize scattering losses and enhance light focusing.

Figure~\ref{fig:opening_angle} explores the impact of emission angle on far-field beam shaping and mode-specific outcoupling performance of the optimized grating coupler.
In Figure~\ref{fig:opening_angle}(a), the far-field intensity profiles along the transverse Y-axis are shown for emission angles ranging from 14$^\circ$ to 18$^\circ$. An increase in angle leads to improved beam confinement and enhanced side-lobe suppression, with the 17$^\circ$ profile achieving over 30\,dB suppression relative to the main lobe. This indicates a stronger directional emission and reduced power leakage into undesired spatial channels.
Figure~\ref{fig:opening_angle}(a) magnifies the beam size relation to the opening angle of the grating. The results confirm that larger angles yield narrower beam widths, which is beneficial for precision light delivery in applications such as individual ion addressing.
Figure~\ref{fig:opening_angle}(c-d) show the outcoupled intensity distributions at a height of $Z$ = $6$$\mu$m above the grating for the $TE_{\text{10}}$ and $TE_{\text{20}}$ waveguide modes, respectively. The $TE_{\text{10}}$ mode produces a single-lobed, symmetric beam profile, while the $TE_{\text{20}}$ mode exhibits a distinct double-lobed pattern, consistent with its modal symmetry. These results demonstrate the grating’s capacity to preserve modal information and enable mode-selective free-space coupling.

\section{$TE_{\text{10}}$ - $TE_{\text{20}}$ Mode Converter}
The adiabatic directional coupler (ADC) was designed to enable efficient mode conversion from $TE_{\text{10}}$ to $TE_{\text{20}}$ at a wavelength of 729 nm \cite{Zhang2014, liu2011,shu2019}. Coupling occurs because the modal field in one waveguide overlaps with the dielectric medium of the other, allowing power to transfer between them. According to coupled mode theory \cite{CMT2021}, the power transfer between two waveguides depends on the coupling coefficient, which is related to the overlap between the mode fields of the two waveguides and the waveguide separation. The power transfer is maximized when the two modes are nearly synchronous, meaning their propagation velocities along the waveguides are comparable. Coupled mode theory describes the evolution of the field amplitudes in both waveguides as a function of the coupling length (z-axis), with the coupled modes exchanging power in a sinusoidal manner. The coupling efficiency depends on the coupling coefficient, which varies with waveguide geometry, refractive index contrast, and separation distance. In the case of adiabatic coupling, the coupling coefficient is engineered to vary gradually along the coupling length to minimize insertion loss and enhance mode conversion \cite{mojaver2024}.
\begin{figure*}[ht]
    \begin{minipage}{1\linewidth}
        \centering
        \includegraphics[width=\linewidth]{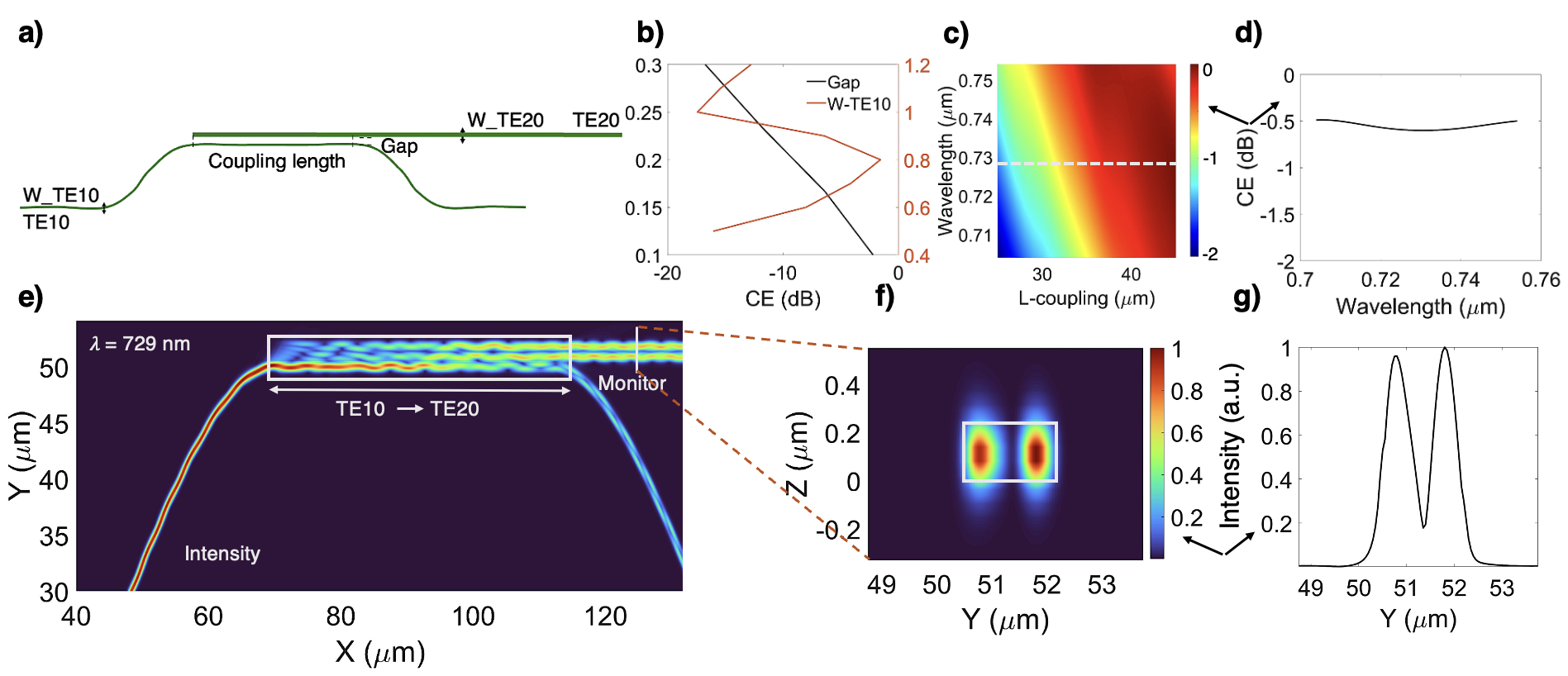}
        \justifying
        \caption{Mode conversion and coupling characteristics of a $TE_{\text{10}}$ to $TE_{\text{20}}$ mode converter. (a) Schematic representation of mode converter with all adjustable parameters. (b) Coupling efficiency based on gap and width of fundamental mode waveguide. (c) Coupling efficiency as a function of coupling length, indicating the optimal length for maximum conversion at 729 nm. (d) Coupling efficiency across a range of wavelengths, illustrating broadband performance. (e) Simulated intensity distribution along the propagation direction, demonstrating efficient conversion from $TE_{\text{10}}$ to $TE_{\text{20}}$. The inset highlights the conversion region. (f)  Cross-sectional mode profile at the output, showing a well-confined $TE_{\text{20}}$ mode. (g) Intensity profile along the Y-axis at the output.}
        \label{fig:ADC}
    \end{minipage}
\end{figure*}
Figure~\ref{fig:ADC}(a) provides a schematic of the ADC geometry, emphasizing the smooth waveguide width transitions that facilitate adiabatic mode evolution and minimize insertion loss. Figure~\ref{fig:ADC}(b) further explores the influence of waveguide width and gap spacing on coupling efficiency, yielding the optimized waveguide widths of 0.8\,\textmu{}m and 1.7\,\textmu{}m for $TE_{\text{10}}$ and $TE_{\text{20}}$, respectively. Figure~\ref{fig:ADC}(c) presents a heatmap of the coupling efficiency across various wavelengths and coupling lengths, with a high-efficiency region near 729 nm that exhibits minimal sensitivity to wavelength fluctuations, indicating robust performance and tolerance to fabrication variations. The broadband coupling behavior shown in Figure~\ref{fig:ADC}(d) demonstrates stable -0.5\,dB performance across a narrow wavelength range, ensuring reliability for practical applications. Figure~\ref{fig:ADC}(e) illustrates the simulated field intensity along the coupling region, demonstrating the gradual transition from $TE_{\text{10}}$ to $TE_{\text{20}}$ over the optimized coupling length of $46$ $\mu$m, with a waveguide gap of 100 nm. The cross-sectional mode profile at the center of the coupling region, shown in Figure~\ref{fig:ADC}(f), highlights the distinct double-lobe characteristic of the $TE_{\text{20}}$ mode. The corresponding intensity distribution along the Y-axis in Figure~\ref{fig:ADC}(g) confirms successful mode conversion, exhibiting the anticipated two-lobe profile. Coupling between the modes allows power to flow from one waveguide to another as a function of the z-axis, with the extent of power transfer dependent on the coupling length. According to couple mode theory, this power transfer is governed by the coupling coefficient, and the conversion efficiency is maximized when the two modes are nearly synchronous in terms of their propagation constants.

\section{Mode Division Multiplexing}
\begin{figure}[ht]
\centering
\includegraphics[width=0.95\linewidth]{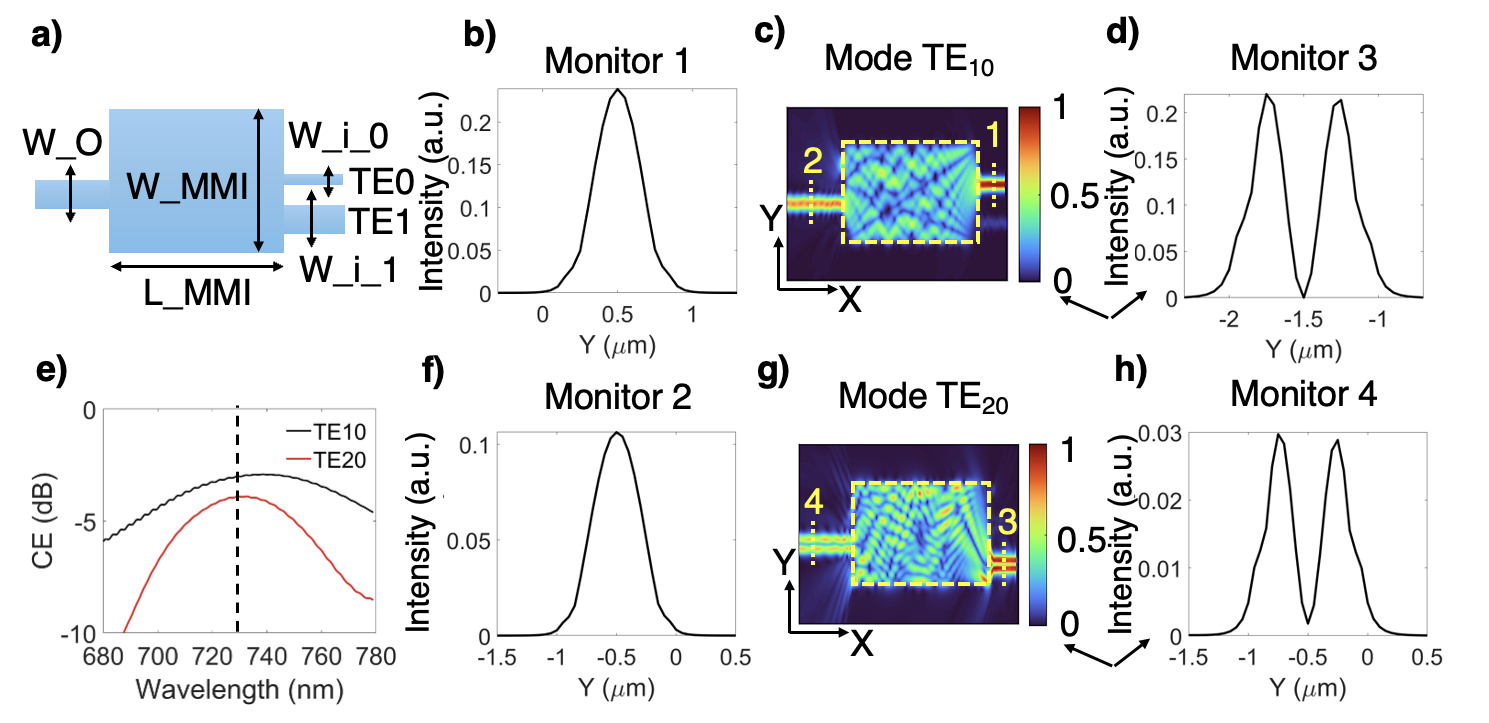}
\caption{Characterization of mode mixing and propagation in a multimode interference (MMI) device. (a) Schematic diagram of the MMI device, illustrating the input waveguide ($w_{\text{io}}$) and output waveguides ($w_{\text{o}}$) for $TE_{\text{10}}$, $TE_{\text{20}}$ modes. The device dimensions ($w_{\text{MMI}}$, $L_{\text{MMI}}$) are designed for optimal mode mixing/selecting and minimal crosstalk. (b, d, f, h) Intensity profiles along the Y-axis at different monitor positions, capturing mode distributions. (c, g) Electric field intensity distributions for $TE_{\text{10}}$ and $TE_{\text{20}}$ modes at the output, highlighting distinct mode patterns. (e)  Coupling efficiency (CE) of $TE_{\text{10}}$ to $TE_{\text{20}}$ modes as a function of wavelength, showing efficient mode selection around the target wavelength.}
\label{fig:MMI}
\end{figure}
The design of the mode division multiplexer (MDM) – multi-mode interferometer (MMI) 2$\times$1 for $TE_{\text{10}}$ and $TE_{\text{20}}$ mode mixing was carefully optimized to ensure efficient coupling and separation of modes, with the objective that if only one input mode is excited (e.g., $TE_{\text{10}}$), the corresponding output remains $TE_{\text{10}}$, and similarly for $TE_{\text{20}}$. The waveguide widths were optimized to 0.6\,\textmu{}m for $TE_{\text{10}}$, 0.9\,\textmu{}m for $TE_{\text{20}}$, and an output waveguide width of 0.9\,\textmu{}m. The MMI width is designed to be 5\,\textmu{}m, with a coupling length of 32\,\textmu{}m, to facilitate the required mode conversion. The device demonstrates a coupling efficiency of -4.6\,dB for $TE_{\text{10}}$ and -4.8\,dB for $TE_{\text{20}}$. The reduced efficiency for $TE_{\text{20}}$ can be attributed to several factors, 
including a suboptimal mode field overlap between $TE_{\text{20}}$ and the surrounding dielectric medium, as well as the inherent phase mismatch between the modes. These factors contribute to a lower coupling efficiency for $TE_{\text{20}}$, despite the careful optimization of the MMI length and waveguide dimensions. Moreover, $TE_{\text{20}}$ mode propagation is more sensitive to variations in waveguide geometry and refractive index contrast, further limiting its coupling efficiency. The optimized design provides a balance between coupling performance and insertion loss, with the MMI structure engineered to ensure minimal distortion and robust mode mixing, while maintaining a high degree of fidelity in mode transmission.




\end{appendices}

\end{document}